\documentclass[preprint,showpacs,preprintnumbers,amsmath,amssymb,floatfix,endfloats*]{revtex4}

\usepackage{graphicx}
\usepackage{dcolumn}
\usepackage{bm}
\usepackage{amsfonts}
\DeclareMathAlphabet{\mathsfsl}{OT1}{cmr}{bx}{it}
\begin{document}

\title{The effective slip length and vortex formation in laminar flow over a rough surface}

\author{Anoosheh Niavarani and Nikolai V. Priezjev}

\affiliation{Department of Mechanical Engineering, Michigan State
University, East Lansing, Michigan 48824}

\date{\today}

\begin{abstract}

The flow of viscous incompressible fluid over a periodically
corrugated surface is investigated numerically by solving the
Navier-Stokes equation with the local slip and no-slip boundary
conditions. We consider the effective slip length which is defined
with respect to the level of the mean height of the surface
roughness. With increasing corrugation amplitude the effective
no-slip boundary plane is shifted towards the bulk of the fluid,
which implies a negative effective slip length. The analysis of the
wall shear stress indicates that a flow circulation is developed in
the grooves of the rough surface provided that the local boundary
condition is no-slip. By applying a local slip boundary condition,
the center of the vortex is displaced towards the bottom the grooves
and the effective slip length increases. When the intrinsic slip
length is larger than the corrugation amplitude, the flow
streamlines near the surface are deformed to follow the boundary
curvature, the vortex vanishes, and the effective slip length
saturates to a constant value. Inertial effects promote vortex flow
formation in the grooves and reduce the effective slip length.

\end{abstract}




\maketitle

\section{Introduction}
\label{sec:Introduction}

An accurate flow prediction in micro-channels is important for the
optimal design and fabrication of microfluidic devices whose
applications range from medicine to
biotechnology~\cite{Walker02,Squires05}. The boundary conditions and
the surface topology are major factors affecting the flow pattern
near the solid boundary and the formation of recirculation zones.
The flow separation at rough surfaces can modify the wall shear
stress distribution or initiate instability towards turbulence. In
microfluidic channels, the vortex flow enhances the mixing
efficiency~\cite{Girault02,Pushpavanam08} and promotes convective
heat transfer~\cite{Manglik04,Haynes06,Haynes07}. In cardiovascular
systems, the separation region at the entrance of branching vessels
may trap lipid particles which could lead to arterial
diseases~\cite{Young70,Wooton99,Berger00}. In the present study we
examine the role of slip boundary condition in determining the flow
properties near rough surfaces including the separation phenomena,
and distribution of pressure and shear stress along the surface.

Although the validity of the no-slip boundary condition is well
accepted at the macroscopic level, recent
experiments~\cite{Churaev84,Vinograd00,Leger00,Breuer03,Charlaix05,Vinograd06}
and molecular dynamics (MD)
simulations~\cite{Thompson90,Nature97,Barrat99,Quirke01,Priezjev04,Priezjev07}
reported the existence of a boundary slip in microflows. The model
first proposed by Navier relates the slip velocity to the rate of
shear via the proportionality coefficient, the so-called slip
length. The MD simulations are particularly suitable for examining
the influence of molecular parameters on the microscopic slip length
at the liquid/solid interface. The advantage of the MD simulations
is that a detailed flow analysis can be performed at the molecular
level while the explicit specification of the boundary conditions is
not required. In contrast to description of the flow near boundary
by means of microscopic slip length, it is convenient to
characterize the flow over macroscopically rough surfaces by the
effective slip length, which is defined as a distance from the level
of the mean height of the surface roughness to the point where
linearly extrapolated bulk velocity profile vanishes. Recent MD
studies have demonstrated that the effective slip length in a flow
of simple fluids~\cite{Priezjev06} and polymer
melts~\cite{Anoosheh08} over a wavy surface agrees well with
hydrodynamic predictions~\cite{Einzel90,Einzel92} when the
corrugation wavelength is larger than approximately thirty molecular
diameters.

The influence of surface roughness on fluid flow with either local
no-slip or zero shear stress (i.e. perfect slip) boundary conditions
has been extensively studied in the past
decades~\cite{Richardson73,Hocking76,Janson88,Miksis94,Tuck95,Sarkar96,Wang03},
see also a review section in~\cite{Priezjev06}. Analytical
calculations have shown that in a shear flow over a corrugated
surface with microscopic no-slip or zero shear stress conditions,
the effective boundary slip is insignificant
macroscopically~\cite{Richardson73,Janson88}. The effective no-slip
boundary plane is located at the intermediate position between
crests and valleys of the rough surface when the no-slip condition
is imposed along the solid
boundary~\cite{Hocking76,Tuck95,Feuillebois04,Vinograd06,HartingPRL07}.
For an arbitrary surface roughness with small amplitudes, the slip
coefficient in the Navier model is proportional to the average
amplitude of the wall roughness and depends on the position of the
origin of the coordinate system~\cite{Miksis94}. Applying the
no-slip boundary condition along the wavy surface, Tuck and
Kouzoubov~\cite{Tuck95} have demonstrated that the effective slip
length is inversely proportional to the corrugation wavelength and
quadratically proportional to the amplitude of the surface
roughness. However, the series expansion method used
in~\cite{Tuck95} fails at large wavenumbers, $ka\gtrsim0.5$, when a
backflow appears inside the grooves of the substrate. The effective
slip length for a flow above the surface with deep corrugations only
weakly depends on the depth of the
grooves~\cite{Hocking76,Wang03,Anoosheh08}. Despite considerable
analytical efforts, the relation between the vortex flow structure
in deep grooves and the effective slip length has not yet been
systematically investigated.

The laminar flow separation at the corrugated surface with the local
no-slip boundary conditions depends on the depth of the grooves and
the Reynolds
number~\cite{Chow72,Bordner78,Taneda79,Sobey80,Nishimura84,Tsangaris86,Leneweit99,Khayat03}.
In a creeping flow over a sinusoidal surface, the flow circulation
appears in sufficiently deep grooves and, as the corrugation
amplitude increases, the vortex grows and remains
symmetric~\cite{Tsangaris86,Pozrikidis87,Aksel04}. With increasing
Reynolds number, the vortex flow forms even in shallow grooves, the
circulation region expands, and the center of vorticity is displaced
upstream~\cite{Chow72,Sobey80,Nishimura84,Leneweit99}. In the limit
of small-scale surface roughness and for no-slip boundary
conditions, the apparent slip velocity at the mean surface becomes
more negative as the Reynolds number increases~\cite{Tuck95}. A
noticeable change in the effective slip length was also observed at
$Re\gtrsim100$ for laminar flow over deep grooves when the local
slip length is comparable to the corrugation
amplitude~\cite{Anoosheh08}. However, the influence of the local
slip condition at the curved boundary on the vortex flow formation
has not been considered at finite Reynolds numbers.


This paper is focused on investigation of the effects of local slip
boundary conditions and the Reynolds number on the flow structure
near periodically corrugated surfaces and the effective slip length.
We will show that for the Stokes flow with the local no-slip
boundary condition, the effective slip length decreases with
increasing corrugation amplitude and the flow circulation develops
in sufficiently deep grooves. In the presence of the local slip
boundary condition along the rough surface, the effective slip
length increases and the size of the vortex is reduced but its
structure remains symmetrical. The analysis of the numerical
solution of the Navier-Stokes equation with the local slip condition
shows that the inertial effects promote the asymmetric vortex flow
formation and reduce the effective slip length.

This paper is organized as follows. The details of a continuum model
and the implementation of the local slip boundary conditions are
described in the next section. The analytical results for the Stokes
flow over a wavy surface by Panzer\,\textit{et\,al.}~\cite{Einzel92}
are summarized in Sec.\,\ref{analytic-secA}. The analysis of the
effective slip length and the flow structure is presented in
Sec.\,\ref{noslip-sec} for the no-slip case and in Sec.\,\ref{slip}
for the finite microscopic slip. The effect of Reynolds number on
the effective slip flow over a periodically corrugated surface is
studied in Sec.\,\ref{slip-re}. A brief summary is given in the last
section.

\section{Details of numerical simulations}
\label{sec:1}

\begin{figure}
\begin{center}
\includegraphics[width=3in,angle=0]{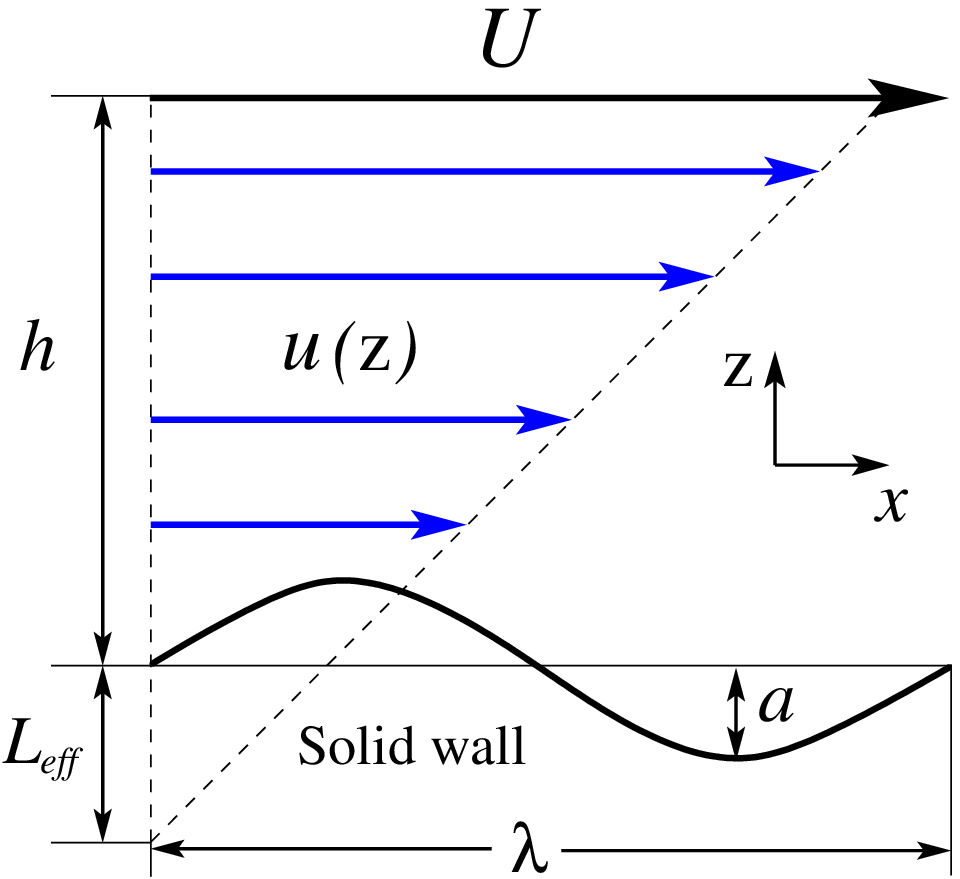}
\caption{(Color online) Schematic diagram of the steady-state
Couette flow over a rough surface. The upper flat wall is moving
with a constant velocity $U$ in the $\hat{x}$ direction. The lower
stationary wall is modeled as a sinusoidal wave with amplitude $a$
and wavelength $\lambda$. The wavenumber $ka\,{=}\,2\,\pi a/\lambda$
varies in the range $0\leqslant ka\leqslant1.26$.} \label{schem}
\end{center}
\end{figure}

\begin{figure}
\begin{center}
\includegraphics[width=3.0in,angle=0]{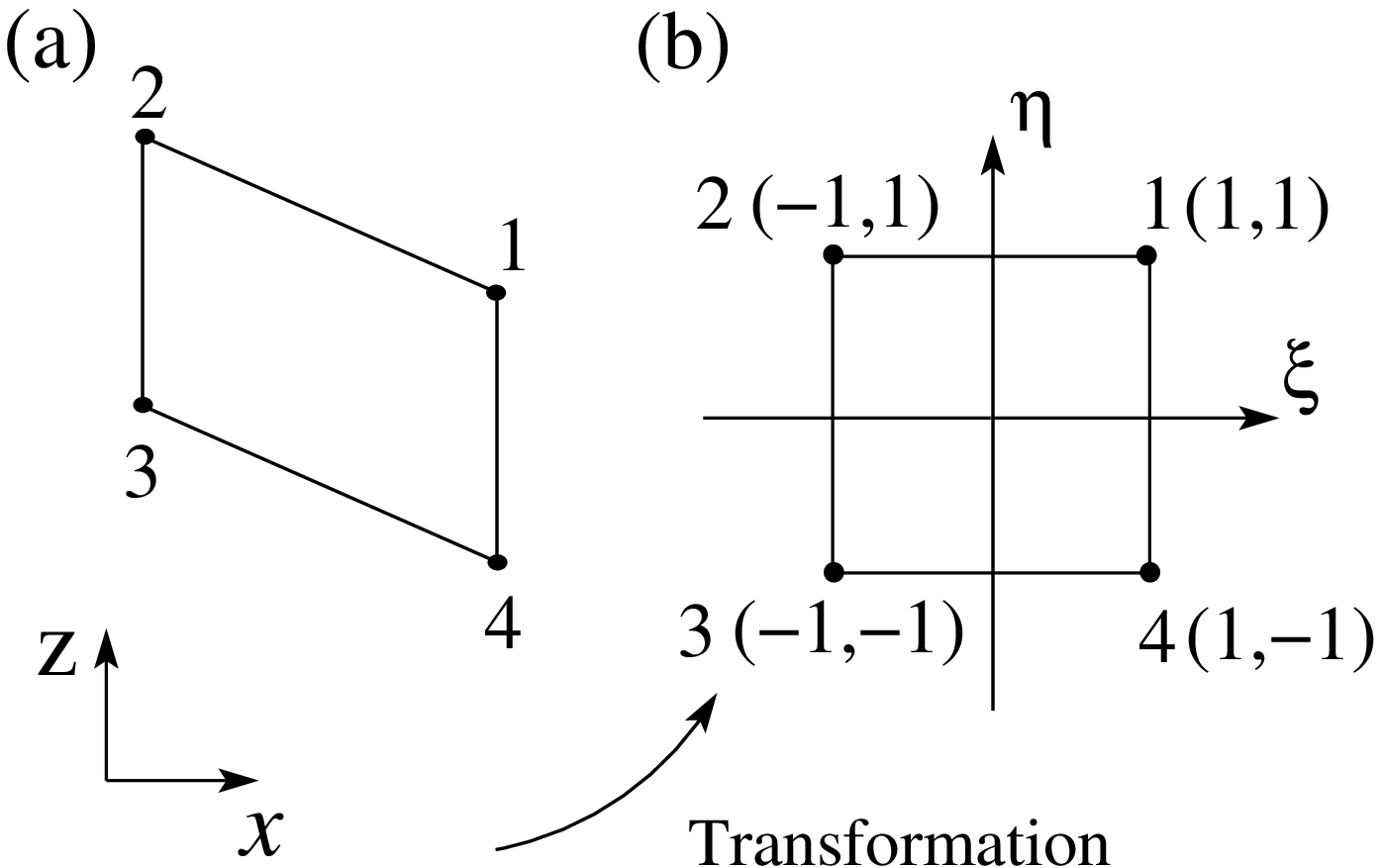}
\caption{Diagram of a bilinear element in (a) the \emph{physical}
coordinate system $(x,z)$ and (b) a transformed element in the
\emph{natural} coordinate system $(\xi,\eta)$.}
\label{coor}
\end{center}
\end{figure}

The two-dimensional incompressible and steady Navier-Stokes (NS)
equation is solved using the finite element method. The
computational setup consists of a viscous fluid confined between an
upper flat wall and a lower sinusoidal wall (see Fig.\,\ref{schem}).
The corrugation wavelength of the lower wall is set to $\lambda$ and
is equal to the system size in the $\hat{x}$ direction. The upper
wall is located at $h=\lambda$ above the reference line at $a=0$,
which is defined as the mean height of the surface roughness. The
continuity and NS equations are
\begin{equation}
\nabla\cdot\textbf{u}=0,
\label{continuuity}
\end{equation}
\begin{equation}
\rho\,(\textbf{u}\cdot\nabla\textbf{u})=-\nabla
p+\mu\nabla^{2}\textbf{u},
\label{N-S}
\end{equation}
where $\textbf{u}=u\,\hat{i}+v\,\hat{\!j}$ is the velocity vector in
the cartesian coordinate system, $p$ is the pressure field, $\rho$
is the fluid density, and $\mu$ is the Newtonian viscosity.

The penalty formulation is employed to avoid decoupling between the
pressure and velocity fields~\cite{Heinrich81}. In this method, the
continuity equation is replaced with a perturbed equation
\begin{equation}
\nabla\cdot\textbf{u}=-\frac{p}{\Lambda},
\label{penalty_formula}
\end{equation}
where $\Lambda$ is the penalty parameter, which ensures the
incompressibility condition. Thus, the modified momentum equations
in the $\hat{x}$ and $\hat{z}$ directions are
\begin{equation}
\rho\,(\textbf{u}\cdot\nabla
u)=\Lambda\nabla(\nabla\cdot\textbf{u})+\mu\nabla^{2} u,
\label{penalty_finalx}
\end{equation}
\begin{equation}
\rho\,(\textbf{u}\cdot\nabla
v)=\Lambda\nabla(\nabla\cdot\textbf{u})+\mu\nabla^{2} v.
\label{penalty_finaly}
\end{equation}
The advantage of the penalty formulation is the elimination of
pressure and the continuity equation. The penalty parameter
$\Lambda$ must be large enough so that compressibility errors are
minimal. The upper bound of $\Lambda$ is determined from the
condition that the viscous effects are not destroyed by the machine
precision~\cite{Heinrich81,Hughes79}. The penalty parameter
$\Lambda$ should be chosen according to the rule
\begin{equation}
\Lambda=c \max\{\mu,\mu Re\},
\end{equation}
where $Re$ is the Reynolds number and the constant $c$ is
recommended to be about $10^{7}$ for computations with
double-precision 64 bit words~\cite{Hughes79}.

The Galerkin formulation of Eq.\,(\ref{penalty_finalx}) and
Eq.\,(\ref{penalty_finaly}) can be explicitly rewritten as
\begin{multline}
\Big{[}\int_{\Omega}\rho N_{i}\Big{(}\bar{u}_{i}u_{j}\frac{\partial
N_{j}}{\partial x}+\bar{v}_{i}u_{j}\frac{\partial N_{j}}{\partial
z}\Big{)}\Big{]}+\Big{[}\int_{\Omega}\Lambda\frac{\partial
N_{i}}{\partial x}\Big{(}\frac{\partial N_{j}}{\partial
x}u_{j}+\frac{\partial N_{j}}{\partial z}v_{j}
\Big{)}d\Omega\Big{]}+\\
\Big{[}\int_{\Omega}\mu\Big{(}\frac{\partial N_{i}}{\partial
x}\frac{\partial N_{j}}{\partial x}+\frac{\partial N_{i}}{\partial
z}\frac{\partial N_{j}}{\partial z}\Big{)}u_{j}
d\Omega\Big{]}=RHS_{x}, \label{galerkinx}
\end{multline}
\begin{multline}
\Big{[}\int_{\Omega}\rho N_{i}\Big{(}\bar{u}_{i}v_{j}\frac{\partial
N_{j}}{\partial x}+\bar{v}_{i}v_{j}\frac{\partial N_{j}}{\partial
z}\Big{)}\Big{]}+\Big{[}\int_{\Omega}\Lambda\frac{\partial
N_{i}}{\partial z}\Big{(}\frac{\partial N_{j}}{\partial
x}u_{j}+\frac{\partial N_{j}}{\partial z}v_{j}
\Big{)}d\Omega\Big{]}+\\
\Big{[}\int_{\Omega}\mu\Big{(}\frac{\partial N_{i}}{\partial
x}\frac{\partial N_{j}}{\partial x}+\frac{\partial N_{i}}{\partial
z}\frac{\partial N_{j}}{\partial z}\Big{)}v_{j}
d\Omega\Big{]}=RHS_{z}, \label{galerkiny}
\end{multline}
where $N_{i}$ is the weight function, $N_{j}$, $u_{j}$, $v_{j}$ are
the node shape function and the velocities in each element,
$\bar{u}_{i}$ and $\bar{v}_{i}$ are the lagged velocities, and the
right hand side ($RHS_{x}$, $RHS_{z}$) terms include the boundary
velocities.

In our simulation, the bilinear quadrilateral elements
$(i,j=1,2,3,4)$ with non-orthogonal edges are transformed to
straight-sided orthogonal elements by introducing the natural
coordinates $\xi=\xi(x,z)$ and $\eta=\eta(x,z)$. The shape functions
$N_i$ in the natural coordinate system are defined as
\begin{equation}
N_{i}=\frac{(1+\xi_{i}\xi)(1+\eta_{i}\eta)}{4} \;\;\;\;\;\;\;
i=1,..,4 \label{shape_function}
\end{equation}
where $\xi_{i}$ and $\eta_{i}$ are the corner points of each element
(see Fig.\,\ref{coor}).

In the next step, Eq.\,(\ref{galerkinx}) and Eq.\,(\ref{galerkiny})
are integrated numerically using four-point Gaussian
quadrature~\cite{Pepper}. The final system of equations is
constructed as follows:
\begin{equation}
\Big{[}\rho [K_{1}] + \Lambda [K_{2}] + \mu [K_{3}]\Big{]}\bigg [
                                                        \begin{array}{c}
                                                          u \\
                                                          v \\
                                                        \end{array}
\bigg]=\bigg [
                                                        \begin{array}{c}
                                                          RHS_{x} \\
                                                          RHS_{z} \\
                                                        \end{array}
                                                        \bigg],
\label{main_NS}
\end{equation}
where the terms $RHS_{x}$ and $RHS_{z}$ contain the velocities at
the boundary nodes.

The boundary conditions must be specified at the inlet, outlet and
upper and lower walls of the Couette cell. The periodic boundary
conditions are imposed at inlet and outlet along the $\hat{x}$
direction. A finite slip is allowed along the lower wall while the
boundary condition at the upper wall is always no-slip. In the local
coordinate system (spanned by the tangential $\vec{t}$ and normal
$\vec{n}$ vectors), the fluid velocity along the lower wavy wall is
computed from
\begin{equation}
u_{t}=L_{0}[(\vec{n}\cdot\nabla)u_{t}+u_{t}/R(x)],
\label{slip_boundary}
\end{equation}
where $u_{t}$ is the tangential component of
$\textbf{u}=u_{t}\vec{t}+u_{n}\vec{n}$, $L_{0}$ is the intrinsic (or
microscopic) slip length at the flat surface, and $R(x)$ is the
local radius of curvature~\cite{Einzel92}. The radius of curvature
is positive for concave and negative for convex regions. The Navier
slip condition for a flat wall is recovered from
Eq.\,(\ref{slip_boundary}) when $R(x)\rightarrow\infty$. The
effective slip length $L_{\textit{eff}}$ at the corrugated lower
wall is obtained by extrapolating the linear part of the velocity
profile ($0.45\leqslant z/h\leqslant0.9$) to zero velocity with
respect to the reference line $a=0$.

The simulation begins by setting the no-slip boundary condition at
the upper and lower walls as an initial guess. Once
Eq.\,(\ref{main_NS}) is solved, the fluid velocities at the lower
boundary are updated using Eq.\,(\ref{slip_boundary}). This
iteration is repeated until the solution is converged to a desired
accuracy. The convergence rate of the solution remains under control
by using the under-relaxation value $0.001$ for the boundary nodes.
The results presented in this paper are obtained with the grid
resolution $150\times150$ in the $\hat{x}$ and $\hat{z}$ directions,
respectively. In order to check the accuracy of the results, several
sets of simulations were also carried out with a finer grid
$180\times180$. The maximum relative error of the effective slip
length due to the grid size is $L_{\textit{eff}}/h=0.003$. The
converged solution of the Navier-Stokes equation satisfies the
following boundary condition:
\begin{equation}
u_{t}=L_{s}(x)\frac{\partial u_{t}}{\partial n}, 
\qquad\frac{1}{L_{s}(x)}=\frac{1}{L_{0}}-\frac{1}{R(x)},
\label{Ls}
\end{equation}
where $L_{s}(x)$ is the local slip length in the presence of surface
curvature~\cite{Einzel92}.

The accuracy of the numerical solution is checked by the normalized
average error, which is defined as
\begin{equation}
error=\Big[\sum^{N_{p}}_{i=1}\frac{\mid
u_{i}^n-u_{i}^{n+1}\mid}{\mid u_{i}^{n+1}\mid}\Big]/N_{p},
\end{equation}
where $N_{p}$ is the total number of computational nodes,
$u^{n}_{i}$ is the velocity at the node $i$ and time step $n$, and
$u^{n+1}_{i}$ is the velocity in the next time step. The typical
value of the error in the converged solution is less than $10^{-9}$.
Throughout the study, the results are presented in the
non-dimensional form. The length scale, shear rate, shear stress,
and velocity are normalized by $h$, $\dot{\gamma}^{*}$,
$\tau_{w}^{*}$, and $U^{*}$, respectively, where $\dot{\gamma}^{*}$
is the shear rate in the case of no-slip boundary condition at the
flat upper and lower walls, and $\tau_{w}^{*}=\mu\dot{\gamma}^{*}$
and $U^{*}\!=h\dot{\gamma}^{*}$.

\section{Results}
\label{sec:2}
\subsection{Analytical solution of the Stokes equation for viscous flow over a wavy wall}
\label{analytic-secA}


The effect of small periodic surface roughness on the effective slip
length has been previously investigated for pressure-driven flows in
a Couette geometry~\cite{Einzel90,Einzel92}. The analytical solution
of the Stokes equation with boundary conditions Eq.\,(\ref{Ls}) at
the wavy wall with amplitude $a$ and wavelength $\lambda$ was
obtained for two limiting cases and small $ka$. For
$L_{0}/\lambda\ll1$ the effective slip length is given by
\begin{equation}
\lim_{kL_{0}\rightarrow0}L_{\textit{eff}}=L_{0}-ka^{2}\omega_{\circ}(ka),
\label{L_eff_limitlow}
\end{equation}
while in the limit of $L_{0}/\lambda\gg1$ it reduces to
\begin{equation}
\lim_{kL_{0}\rightarrow\infty}L_{\textit{eff}}=\Big(\frac{1}{L_{0}}+\frac{k^{3}a^{2}}{\omega_{\infty}(ka)}\Big)^{-1},
\label{L_eff_limitup}
\end{equation}
where the functions $\omega_{\circ}(ka)$ and  $\omega_{\infty}(ka)$
are defined as
\begin{equation}
\omega_{\circ}(ka)=\frac{1-1/4(ka)^{2}+19/64(ka)^{4}+O[(ka)^{6}]}{1+(ka)^{2}-1/2(ka)^{4}+O[(ka)^{6}]},
\label{w1}
\end{equation}
and
\begin{equation}
\omega_{\infty}(ka)=\frac{1-5/4(ka)^{2}+61/64(ka)^{4}+O[(ka)^{6}]}{1+(ka)^{2}-1/2(ka)^{4}+O[(ka)^{6}]}.
\label{w2}
\end{equation}
An approximate analytical expression for the effective slip length
that interpolates between the two bounds Eq.\,(\ref{L_eff_limitlow})
and Eq.\,(\ref{L_eff_limitup}) is given by
\begin{equation}
L_{\textit{eff}}=\frac{L_{0}\omega_{\infty}(ka)-ka^{2}\omega_{0}(ka)/(1+2kL_{0})}{1+k^{3}a^{2}L_{0}},
\label{L_eff}
\end{equation}
with the range of applicability $ka\lesssim0.5$. For larger
wavenumbers $ka>0.5$, the function $\omega_{\infty}(ka)$
overestimates the numerical solution and the interpolated formula
Eq.\,(\ref{L_eff}) does not apply~\cite{Einzel92}.

\subsection{Flow over a rough surface with the local no-slip boundary condition}
\label{noslip-sec}


\begin{figure}[t]
\begin{center}
\includegraphics[width=5in,angle=0]{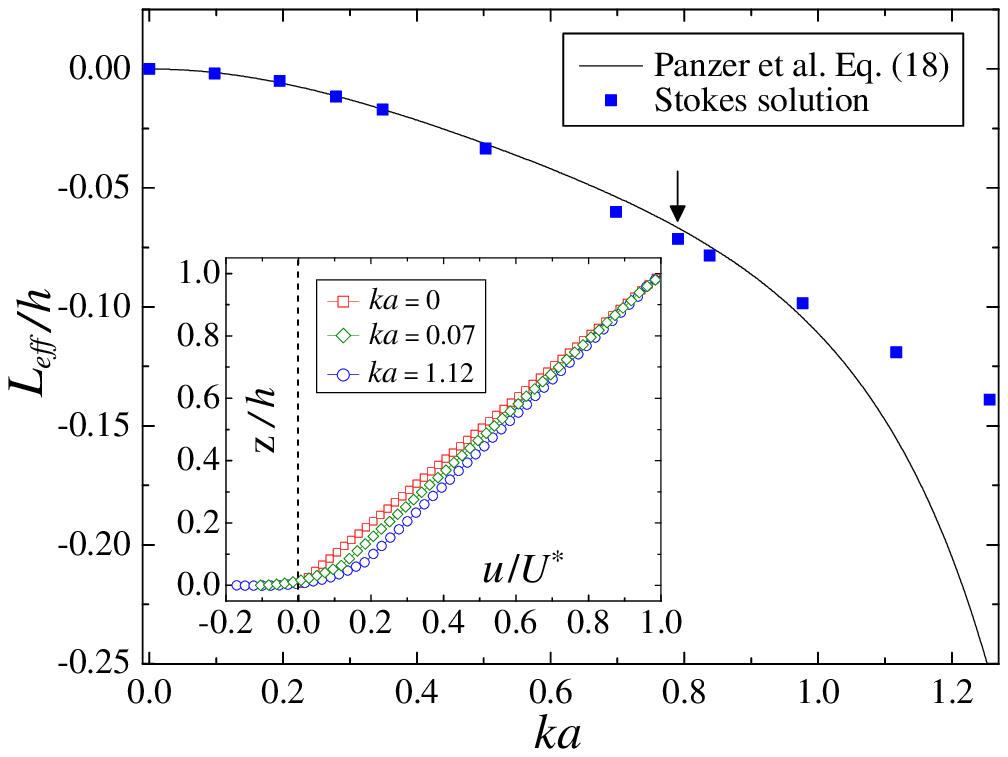}
\caption{(Color online) The effective slip length as a function of
wavenumber $ka$ computed from the solution of the Stokes equation
with the no-slip boundary condition. The primary vortex is formed at
the bottom of the valley for $ka\geqslant0.79$ (see the vertical
arrow at $ka=0.79$). The solid line is calculated using
Eq.\,(\ref{L_eff}). The inset shows the normalized velocity profiles
obtained from the Stokes solution for the selected values of $ka$.
The dashed line located at $a=0$ is the reference for calculating
the effective slip length.} \label{einzelnoslip}
\end{center}
\end{figure}

In this section, the Stokes equation with the local no-slip
condition at the upper and lower walls is solved numerically to
study the effect of corrugation amplitude on the effective slip
length. The velocity profiles, averaged over the period of
corrugation $\lambda$, are plotted in the inset of
Fig.\,\ref{einzelnoslip} for several values of wavenumber
$ka\,{=}\,2\,\pi a/\lambda$. As $ka$ increases, the normalized
velocity profiles remain linear in the bulk region and become curved
near the lower corrugated wall. The linear part of the velocity
profiles is used to compute the effective slip length, which is
plotted as a function of wavenumber in Fig.\,\ref{einzelnoslip}.
With increasing corrugation amplitude of the lower wall, the
effective slip length decays monotonically and becomes negative,
indicating that the effective no-slip boundary is shifted into the
fluid domain. For $ka\lesssim1$, the numerical results agree well
with the analytical solution Eq.\,(\ref{L_eff}) denoted by the solid
line in Fig.\,\ref{einzelnoslip}. The deviation from the analytical
solution becomes significant at larger wavenumbers where the
streamlines extracted from the Stokes solution indicate the presence
of backflow at the bottom of the valley.

\begin{figure}
\begin{center}
\includegraphics[width=5in,angle=0]{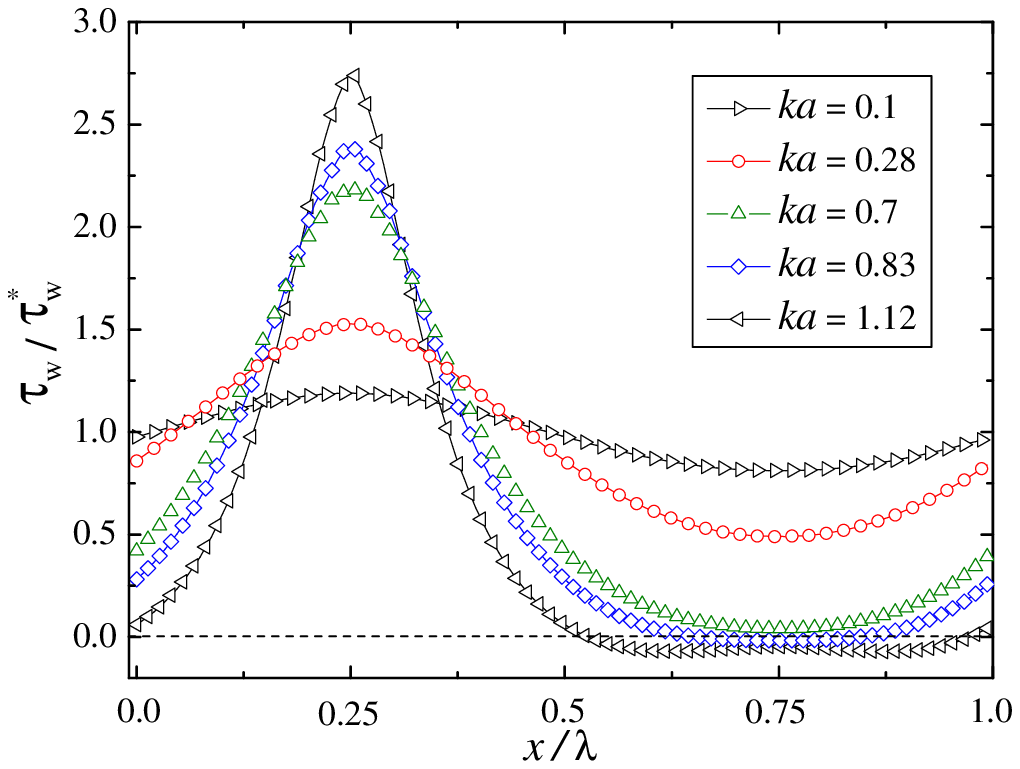}
\caption{(Color online) Shear stress along the lower wavy wall
computed from the Stokes solution with no-slip boundary condition
($L_{0}=0$) for the indicated values of wavenumber $ka$. The value
$\tau_w^{\ast}$ used for normalization is the shear stress at the
flat wall. The intersection of the curves with the dashed line
determines the location of the flow separation and attachment inside
the valley.} \label{shearamp}
\end{center}
\end{figure}

In order to investigate the flow behavior above the sinusoidal
surface, the shear stress and pressure along the lower wall were
computed from the solution of the Stokes equation. In the presence
of surface curvature the wall shear stress $\tau_{w}$ has two
components
\begin{equation}
\tau_{w}=\mu\Big(\frac{\partial u_{t}}{\partial
n}+u_{t}/R(x)\Big)\Big|_{w}, \label{tw}
\end{equation}
where $\partial u_{t}/\partial n$ is the normal derivative of the
tangential velocity $u_{t}$, and $R(x)$ is the local radius of
curvature. In the case of no-slip boundary condition ($u_{t}=0$),
the local shear stress at the wall is reduced to
$\tau_{w}\,{=}\,\,\mu\frac{\partial u_{t}}{\partial n}$. The
normalized shear stress along the lower corrugated wall is plotted
in Fig.\,\ref{shearamp} for different corrugation amplitudes. The
maximum value of the shear stress is located at the peak of the
surface corrugation ($x/\lambda=0.25$) and it increases with
increasing amplitude, which is consistent with the results of
previous analytical studies of a laminar flow over a wavy
wall~\cite{Benjamin59,Bordner78}. The fluid tangential velocity near
the boundary is proportional to the wall shear stress shown in
Fig.\,\ref{shearamp}. Therefore, the tangential velocity is also
maximum above the peak and, as the flow moves downstream, it
decelerates and the velocity becomes zero inside the valley at
sufficiently large amplitudes. For $ka\geqslant0.79$, the shear
stress profiles intersect the dashed line ($\tau_w\!=0$) at two
points and a clockwise flow circulation develops inside the valley.
As the corrugation amplitude increases, the intersection points move
away from each other and the flow recirculation region becomes
larger. These results are in agreement with previous estimates of
the critical wavenumber $ka\thickapprox0.77$ for the onset of flow
separation in sufficiently thick films~\cite{Pozrikidis87,Aksel04}.

\begin{figure}
\begin{center}
\includegraphics[width=5in,angle=0]{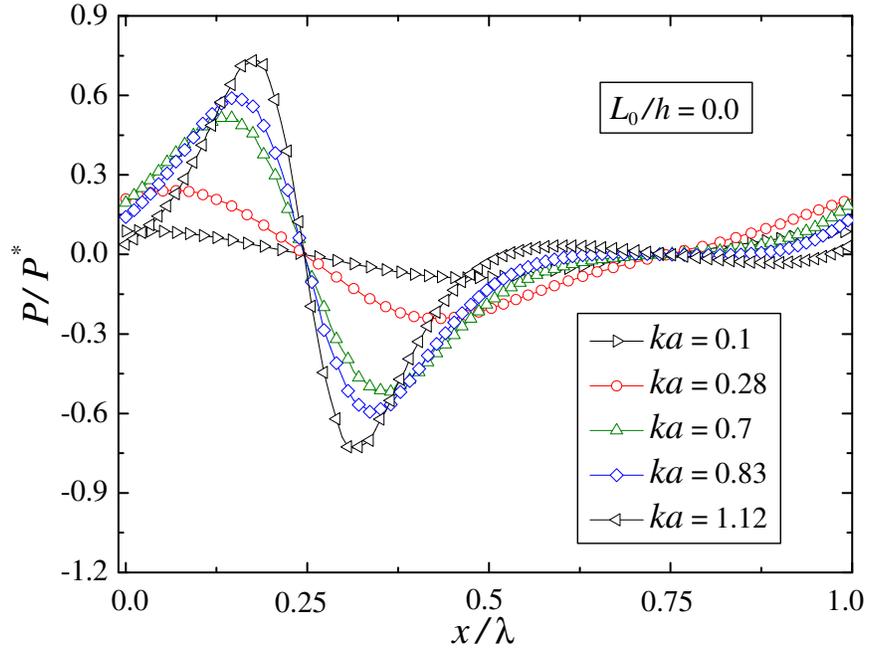}
\caption{(Color online) The normalized pressure along the lower
corrugated boundary extracted from the Stokes solution with no-slip
boundary condition for the same values of $ka$ as in
Fig.\,\ref{shearamp}. The maximum pressure $P^{*}$ for $ka=1.12$ and
$L_{0}=0$ is located in the bulk region at
$(x/\lambda,z/\lambda)\simeq(0.07,0.23)$.} \label{pressure_ka}
\end{center}
\end{figure}

The pressure along the lower wavy wall is plotted in
Fig.\,\ref{pressure_ka} for the same amplitudes as in
Fig.\,\ref{shearamp}. The value $P^{*}$ used for normalization is
the maximum pressure, which is located above the wavy surface with
$ka=1.12$ on the left side of the peak. For each wavenumber, the
pressure along the surface is maximum on the left side of the peak,
where the surface faces the mainstream flow. The surface pressure
reaches its minimum value on the right side of the peak (see
Fig.\,\ref{pressure_ka}). As the flow moves further downstream into
the valley, it encounters an adverse pressure gradient, which
becomes larger as $ka$ increases. At large wavenumbers
$ka\geqslant0.83$, the flow near the surface cannot overcome the
combined resistance of the viscous forces and the adverse pressure
gradient, and separates from the surface at the point where
$\tau_{w}\!=0$.

The pressure contours and streamlines near the corrugated surface
with wavenumber $ka=1.12$ are depicted in
Fig.\,\ref{contourslip}(a). The pressure contours indicate the
presence of an adverse pressure gradient in the region $0.3\lesssim
x/\lambda\lesssim0.6$ on the right side of the peak (see also
Fig.\,\ref{pressure_ka}). The streamlines illustrate the flow
separation inside the valley at $x/\lambda\simeq0.52$. After the
separation point the flow near the wall reverses direction and moves
against the mainstream. The local velocity profile inside the valley
is shown in the inset of Fig.\,\ref{contourslip}(a). As the flow
cross-section decreases towards the peak, the flow attaches to the
surface at the point where the shear stress is zero again. Note also
that for $ka\geqslant0.83$ in Fig.\,\ref{pressure_ka}, the pressure
profiles along the lower surface exhibit a slight drop inside the
valley where the vortex is present. After the attachment point, due
to the periodic boundary condition in the $\hat{x}$ direction, the
surface pressure increases up to its maximum value, which is located
on the left side of the peak. The nonlinearity due to inertia is
absent in the Stokes flow and the vortex in the valley remains
symmetrical.

We also comment that a secondary vortex is formed inside a deeper
cavity ($ka\thickapprox2.33$), which is counter-rotating with
respect to the primary vortex (not shown). Previous analytical study
of a creeping flow over a wavy wall demonstrated that a secondary
vortex appears at $ka\gtrsim2.28$ when the channel width is larger
than the corrugation wavelength~\cite{Aksel04}. In the limiting
case, when the cavity consists of two parallel walls, an infinite
sequence of counter-rotating vortices appears between the
walls~\cite{Pan67,Moffatt64}.

\subsection{Effect of the local slip on the flow pattern near the rough surface}
\label{slip}

\begin{figure}
\begin{center}
  \includegraphics[width=3.0in,angle=0]{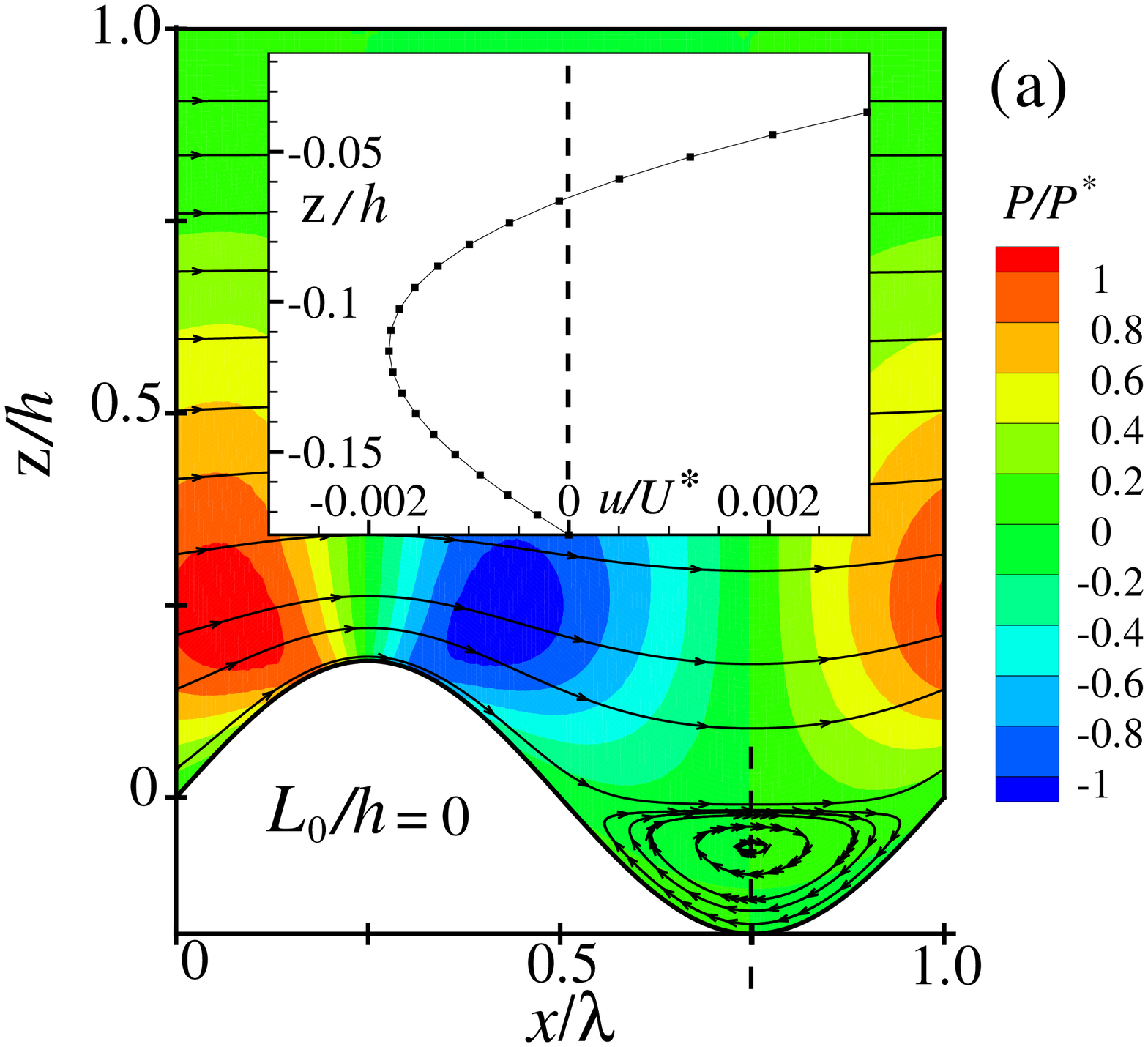}
  \includegraphics[width=3.0in,angle=0]{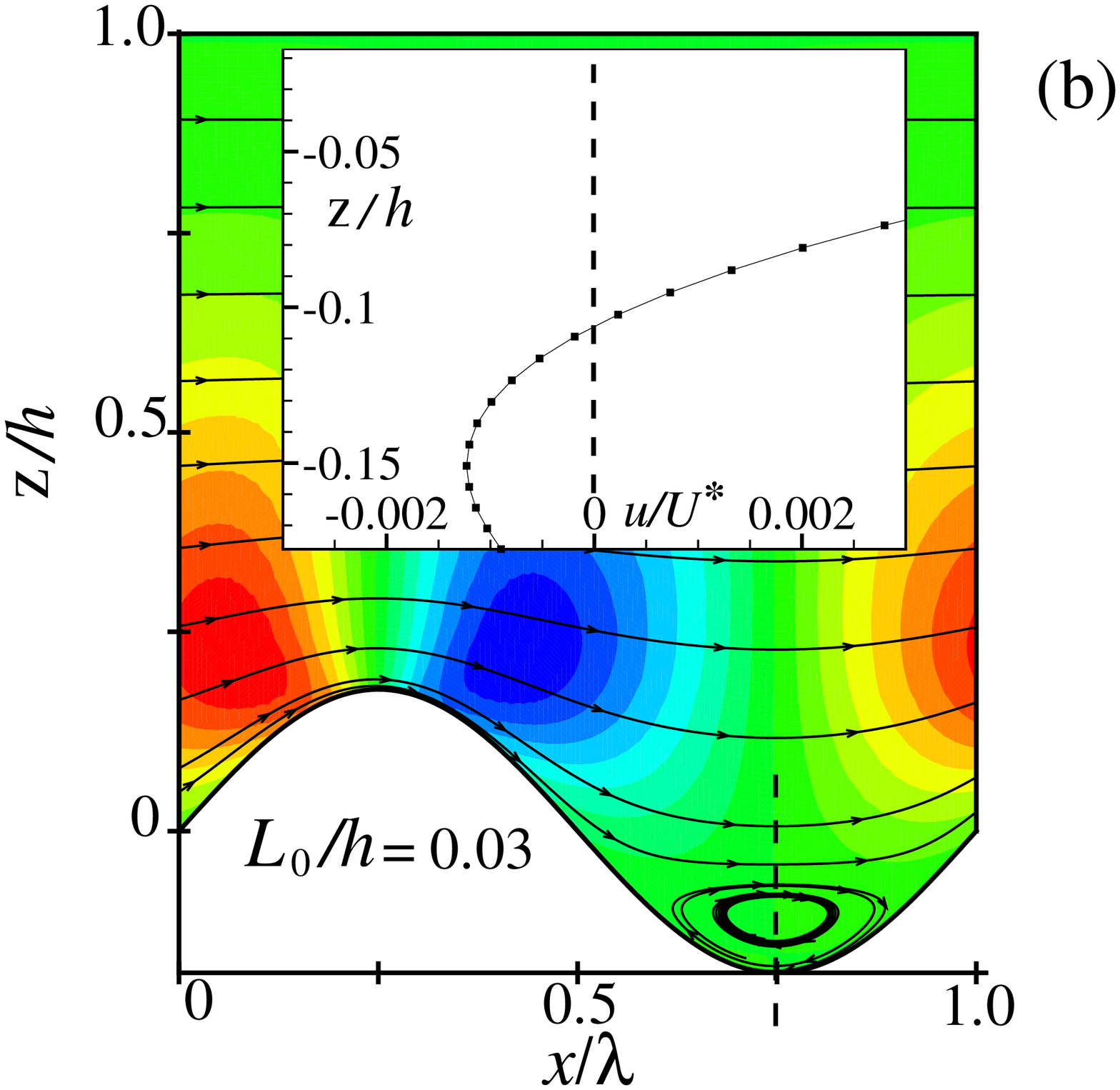}\\
  \includegraphics[width=3.0in,angle=0]{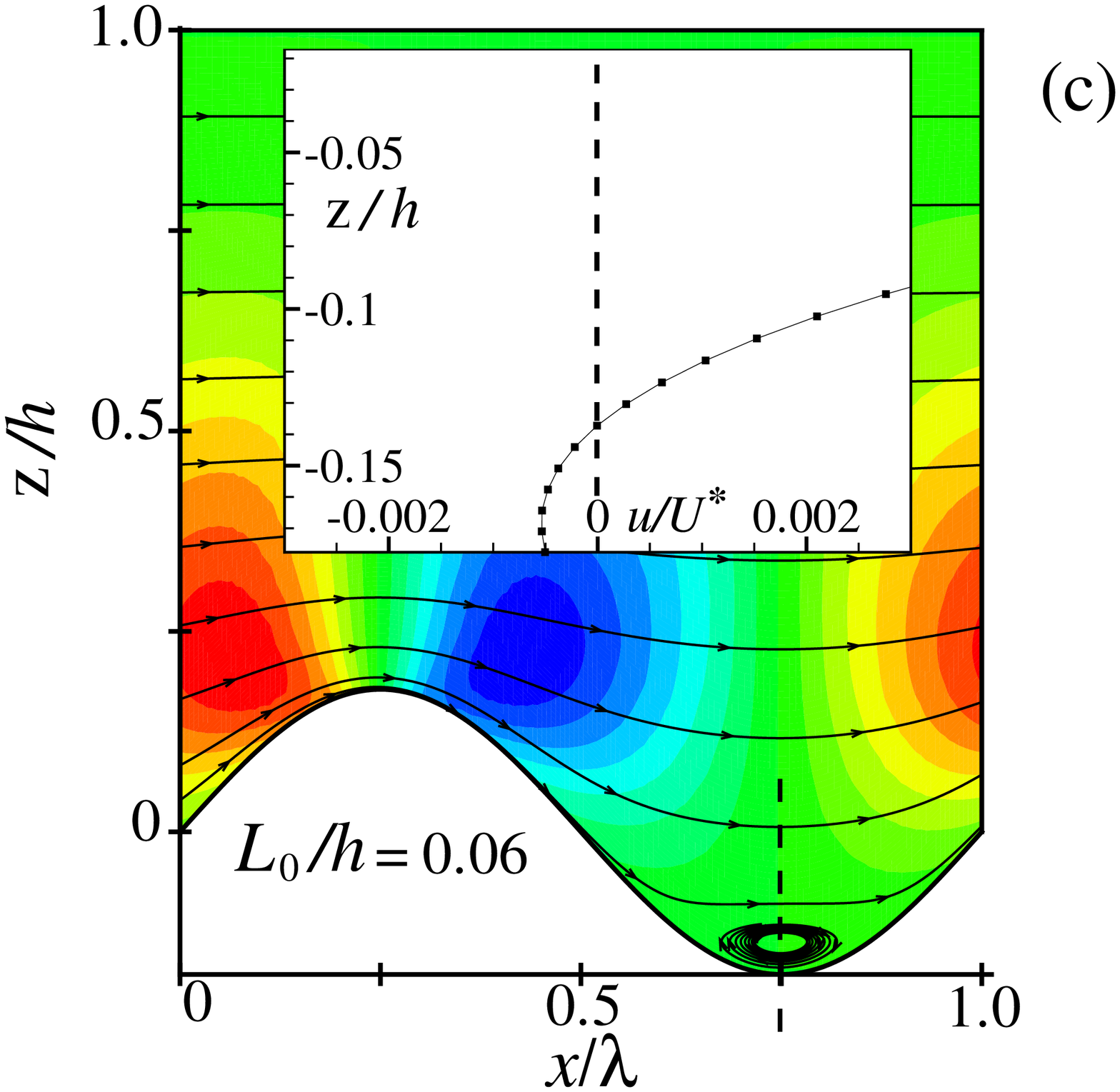}
  \includegraphics[width=3.0in,angle=0]{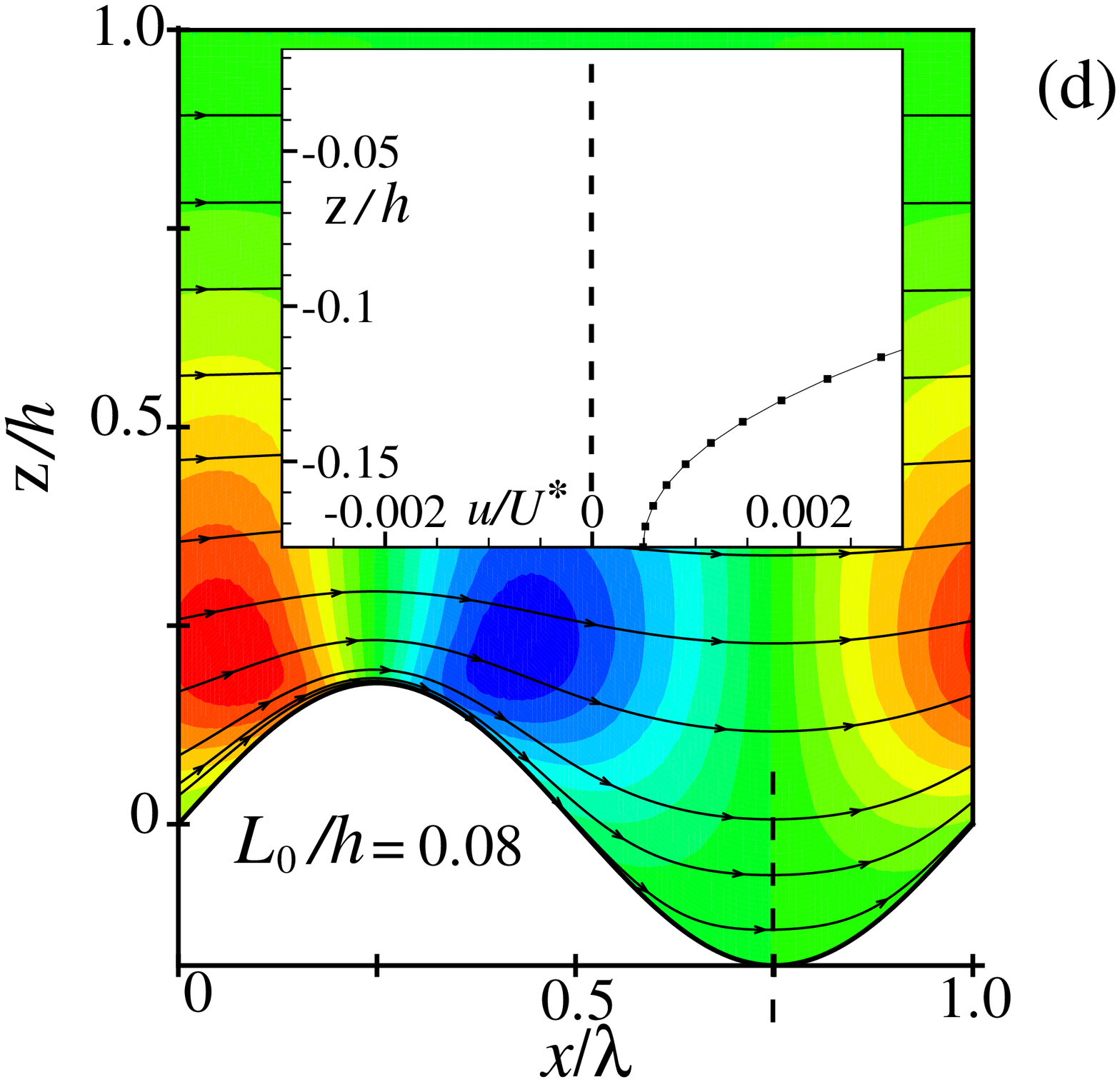}
\caption{(Color online) Pressure contours and streamlines for the
wavenumber $ka=1.12$ and the slip length at the lower wall $L_{0}=0$
(a), $L_{0}/h=0.03$ (b), $L_{0}/h=0.06$ (c), and $L_{0}/h=0.08$ (d).
The pressure contours are normalized by the maximum value $P^{*}$
located on the left side of the peak
$(x/\lambda,z/\lambda)\simeq(0.07,0.23)$ in the case of $L_{0}=0$.
The vertical dashed line inside the valley at $x/\lambda=0.75$
indicates the cross-section used to compute the velocity profiles
shown in the inset.} \label{contourslip}
\end{center}
\end{figure}

Next, we present the results of the numerical solution of the Stokes
equation with the local slip condition at the lower wavy wall while
the boundary condition at the upper wall remains no-slip. The
pressure contours and the streamlines are plotted in
Fig.\,\ref{contourslip} for several values of the intrinsic slip
length $L_{0}$ and the wavenumber $ka=1.12$. The streamline patterns
indicate that with increasing slip length $L_{0}$, the size of
vortex inside the valley is progressively reduced and the vortex
eventually disappears. Similar to the analysis in the previous
section, the pressure and shear stress along the lower wall are
computed in the presence of the local boundary slip.

\begin{figure}
\begin{center}
\includegraphics[width=5in,angle=0]{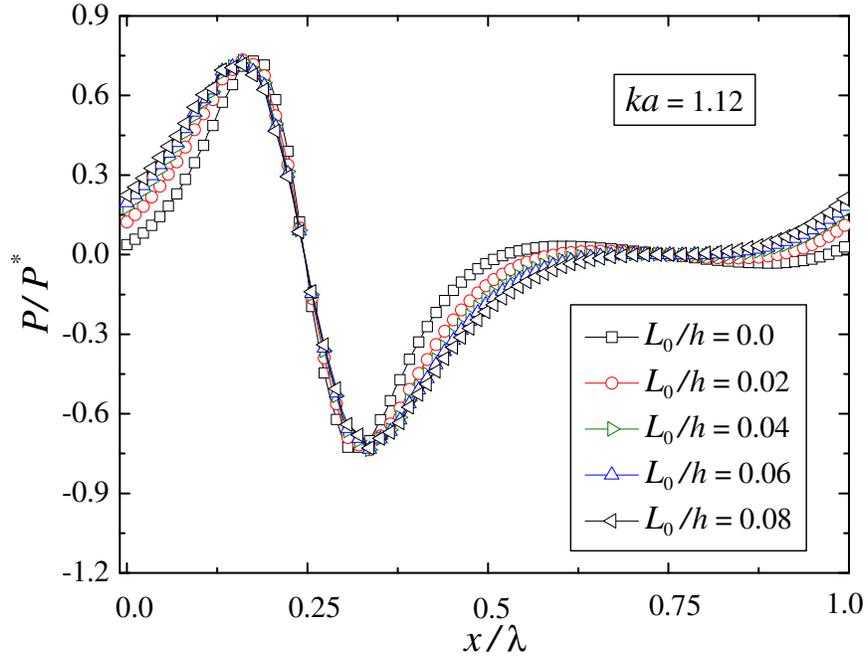}
\caption{(Color online) Normalized pressure along the lower wavy
wall extracted from the solution of the Stokes equation for the
tabulated values of $L_{0}$ and $ka=1.12$. The maximum pressure
$P^{*}$ for $L_{0}=0$ is located above the lower surface at
$(x/\lambda,z/\lambda)\simeq(0.07,0.23)$.} \label{pressure_L0}
\end{center}
\end{figure}

\begin{figure}
\begin{center}
\includegraphics[width=5in,angle=0]{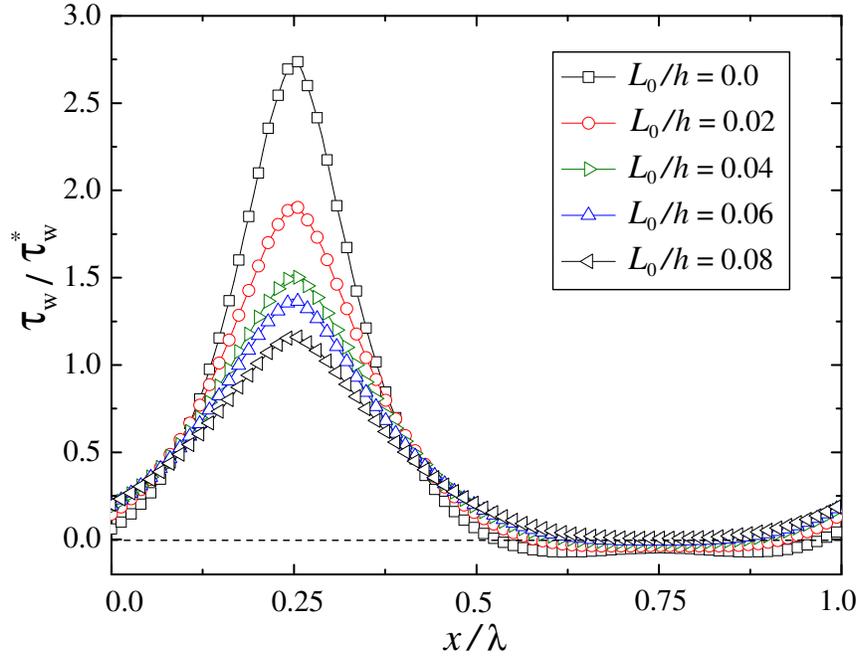}
\caption{(Color online) Shear stress along the corrugated lower wall
($ka=1.12$) for the indicated values of the local slip length
$L_{0}$. The intersection of the dashed line with the shear stress
profiles shows the location of the flow separation and attachment
inside the valley.} \label{shearslip}
\end{center}
\end{figure}

The normalized pressure along the lower wavy wall is plotted in
Fig.\,\ref{pressure_L0} as a function of the slip length $L_{0}$.
Similar to the case of no-slip boundary condition, the profiles
exhibit a maximum and a minimum in pressure on the right and left
sides on the peak, respectively. In a wide range of $L_{0}$, the
difference between the locations and the magnitudes of the extrema
is barely noticeable. As the flow moves down a slope ($0.3\lesssim
x/\lambda\lesssim0.5$ region in Fig.\,\ref{pressure_L0}), the
pressure gradient along the surface becomes positive and its
magnitude decreases with increasing values of $L_{0}$. The
separation and attachment points are located at the intersection of
the shear stress profiles with the horizontal line ($\tau_{w}\!=0$)
shown in Fig.\,\ref{shearslip}. In comparison to the no-slip case,
the smaller combined effect of the adverse pressure gradient and the
wall shear stress causes a shift of the separation point deeper into
the valley (e.g. see Fig.\,\ref{contourslip}). As the slip length
$L_{0}$ increases, the separation and attachment points move closer
to each other, the vortex becomes smaller and eventually vanishes.
We also note that at the separation point both components of the
shear stress $\partial u_{t}/\partial n$ and $u_{t}/R(x)$ become
zero in agreement with the MRS criterion for the flow separation of
the boundary layer at a moving substrate~\cite{Schlichting00}.

The influence of the local slip on the shape of the velocity profile
and the normal derivative of the tangential velocity at the bottom
of the valley is illustrated in the insets of
Fig.\,\ref{contourslip}. The local slip length $L_s(x)$ in
Eq.\,(\ref{Ls}) is a function of the radius of curvature $R(x)$ and
the intrinsic slip length $L_{0}$. For all cases considered in
Fig.\,\ref{contourslip}, the condition $L_0<|R(x)|$ holds and the
local slip length $L_{s}$ remains positive everywhere along the
lower wall, which means that the slip velocity $u_{t}$ and the
normal derivative $\partial u_{t}/\partial n$ carry the same sign.
It is expected, however, that for the opposite condition,
$L_{0}>|R(x)|$, the values $u_{t}$ and $\partial u_{t}/\partial n$
would have different signs at the bottom of the valley, and the
corresponding velocity profile will be qualitatively similar to the
profile shown in the inset of Fig.\,\ref{contourslip}(d) but shifted
by a negative slip velocity (see next section).

\begin{figure}
\begin{center}
\includegraphics[width=5in,angle=0]{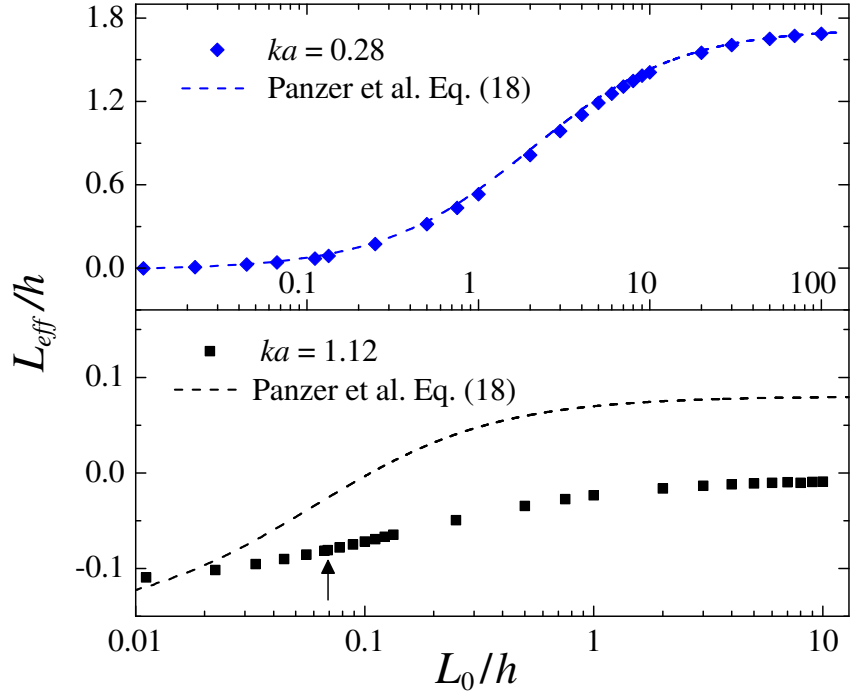}
\caption{The effective slip length, extracted from the Stokes
solution as a function of $L_0$ for $ka=0.28$ (top) and $ka=1.12$
(bottom). The dashed line is computed from Eq.\,(\ref{L_eff}). For
$ka=1.12$ and $L_{0}/h\leqslant0.067$, the vortex is formed in the
valley (see the vertical arrow at $L_{0}/h=0.067$). The error bars
are smaller than the symbol size.} \label{slipadd}
\end{center}
\end{figure}

\begin{figure}
\begin{center}
\includegraphics[width=5in,angle=0]{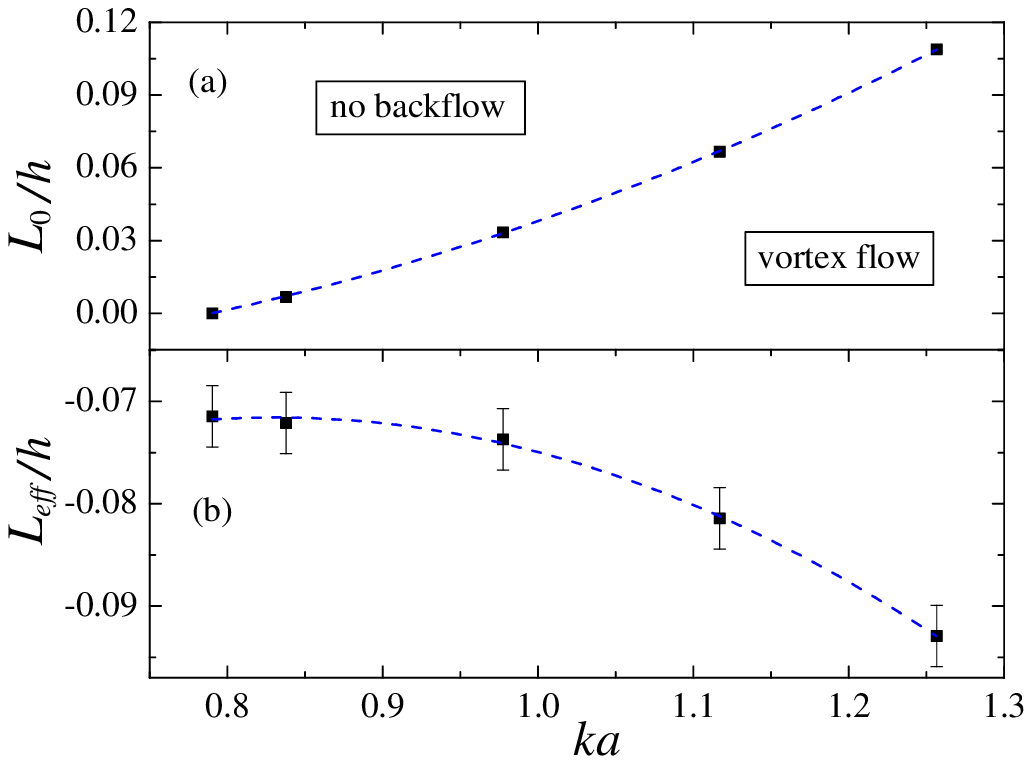}
\caption{(Color online) The intrinsic slip length $L_{0}$ above
which there is no vortex at the bottom of the valley (top) and the
corresponding effective slip length $L_{\textit{eff}}$ (bottom)
computed from the Stokes solution and plotted as a function of
wavenumber $ka$. The increment in the slip length used to determine
the threshold values is about the symbol size (top). The dashed
lines are quadratic fits to guide the eye.} \label{onset_L0}
\end{center}
\end{figure}

The effective slip length computed from the numerical solution of
the Stokes equation and Eq.\,(\ref{L_eff}) is plotted in
Fig.\,\ref{slipadd} as a function of the intrinsic slip length
$L_{0}$ for wavenumbers $ka=0.28$ ($a/h=0.04$) and $ka=1.12$
($a/h=0.18$). For small values of $L_{0}$, the effective slip length
approaches a negative value previously reported in
Fig.\,\ref{einzelnoslip} for the no-slip case. As $L_{0}$ increases,
$L_{\textit{eff}}$ grows monotonically and appears to saturate to a
constant value. The transition of the effective slip length from a
growing function of $L_{0}$ to a nearly constant value occurs at
larger $L_{0}$ when $ka$ decreases. For the small wavenumber
$ka=0.28$, the effective slip length computed from
Eq.\,(\ref{L_eff}) is in a good agreement with the numerical
solution of the Stokes equation. Visual inspection of the streamline
patterns indicates that there is no backflow at any $L_{0}$. In the
saturation regime, $L_{0}\!\rightarrow\!\infty$, the wall shear
stress becomes zero everywhere along the lower boundary, the
streamlines near the lower wall follow the boundary curvature, and
the effective slip length in Eq.\,(\ref{L_eff_limitup}) approaches
$L_{\textit{eff}}/\lambda\simeq1/\,[2\pi(ka)^{2}]-9/8\pi$. For the
large wavenumber $ka=1.12$, the analytical results
Eq.\,(\ref{L_eff}) overestimate $L_{\textit{eff}}$  computed from
the numerical solution at $L_{0}/h\gtrsim0.02$ and the flow
circulation is developed in the valley at $L_{0}/h\leqslant0.067$.
The vortex vanishes at the bottom of the valley at sufficiently
large values of $L_{0}$ (denoted by the vertical arrow in
Fig.\,\ref{slipadd}), and the flow streamlines are deformed to
follow the boundary curvature [e.g. see Fig.\,\ref{contourslip}(d)].
The results for the intrinsic slip length, which determines the
threshold for the onset of the flow circulation at the bottom of the
groove, are summarized in Fig.\,\ref{onset_L0}. For the wavenumbers
examined in this study, $ka\leqslant1.26$, the numerical simulations
indicate that if the flow circulation is present in the valley then
the effective slip length is negative and $L_{\textit{eff}}$
increases with decreasing vortex size.

\subsection{Effect of the Reynolds number on the effective slip length}
\label{slip-re}


The analysis of the Stokes equation discussed in the previous
section demonstrated that increasingly large local slip at the lower
wavy wall eliminates the flow circulation in the valley and leads to
a larger effective slip. In this section, the influence of the
inertia term in the Navier-Stokes equation on the flow pattern and
the effective slip length is investigated. For the shear flow with
slip condition at the lower corrugated wall $L_0$, the Reynolds
number is defined as
\begin{equation}
Re=\frac{\rho\,U^{*}h\,(1+L_{\textit{eff}}/h)}{\mu},
\label{Re_function}
\end{equation}
where $\rho$ is the fluid density, $U^{*}$ is the upper wall
velocity, and $h\,(1+L_{\textit{eff}}/h)$ is the distance between
the upper flat wall and the effective no-slip boundary plane. In the
case of no-slip boundary condition at the lower flat wall,
$L_{\textit{eff}}$ is zero and the standard definition of the
Reynolds number is recovered, i.e., $Re=\rho\,U^{*}h/\mu$.

\begin{figure}
\begin{center}
\includegraphics[width=5in,angle=0]{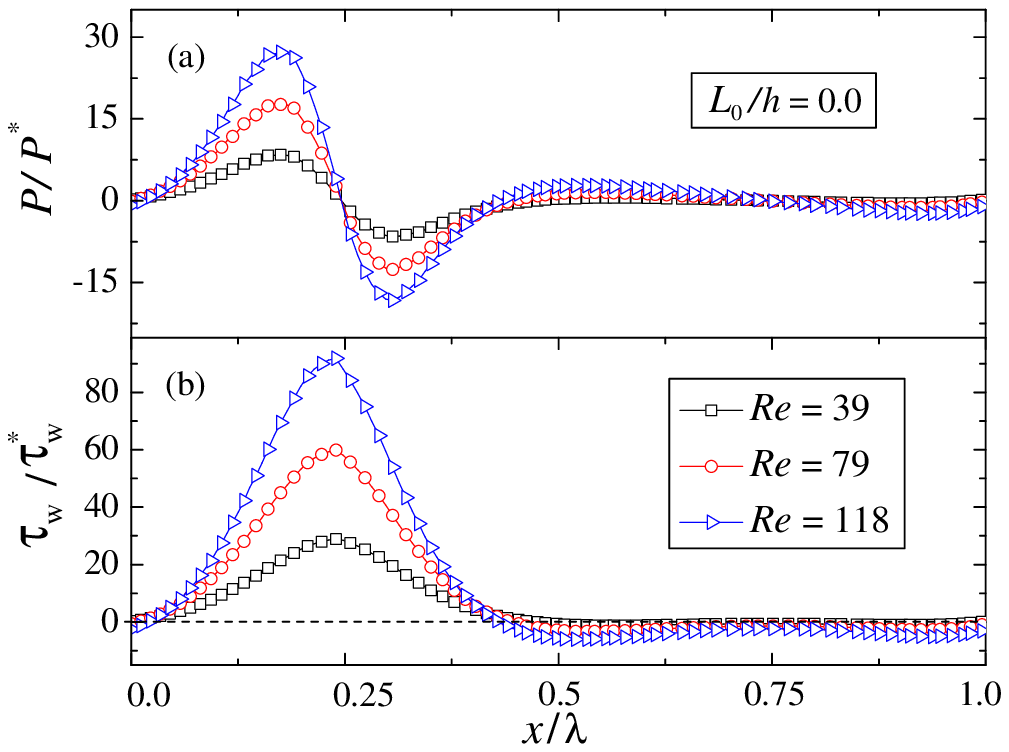}
\caption{(Color online) The normalized pressure (top) and shear
stress (bottom) along the lower wavy wall as a function of the
Reynolds number for $ka=1.12$ and no-slip boundary conditions. The
value $P^{*}$ is the maximum pressure located above the lower
boundary on the left side of the peak
$(x/\lambda,z/\lambda)\simeq(0.07,0.23)$ for $L_{0}=0$ and
$Re=4.0$.} \label{re_noslip}
\end{center}
\end{figure}

The pressure and shear stress along the lower wavy wall are plotted
in Fig.\,\ref{re_noslip} for the selected values of the Reynolds
number and no-slip boundary conditions. As $Re$ increases, the
adverse pressure gradient along the right side of the corrugation
peak ($0.3\lesssim x/\lambda\lesssim0.5$) becomes larger and the
pressure drop inside the valley ($0.5\lesssim x/\lambda\lesssim0.9$)
increases. For each value of the Reynolds number in
Fig.\,\ref{re_noslip}(b), similar to the Stokes flow case, the shear
stress above the peak is maximum and, as the flow moves downstream
along the right side of the peak, it decelerates and eventually
separates from the surface when $\tau_{w}=0$. With increasing upper
wall velocity, the shear stress above the peak increases and causes
the flow to decelerate faster along the right side of the peak. The
separation and attachment points (determined from the condition
$\tau_{w}=0$) move further apart from each other and the circulation
region inside the valley expands [see Fig.\,\ref{re_noslip}(b)].

\begin{figure}
\begin{center}
\includegraphics[width=3.0in,angle=0]{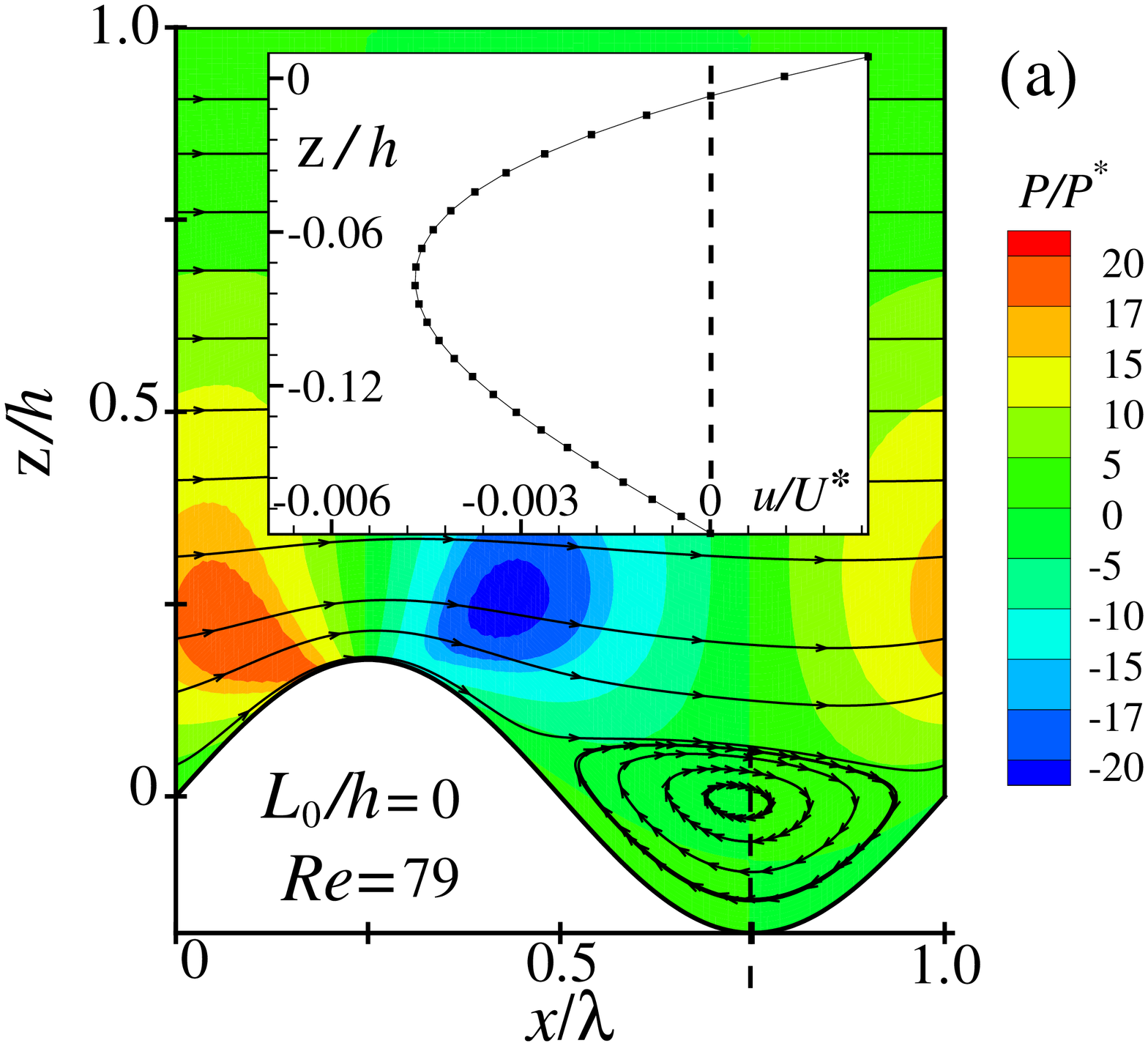}
\includegraphics[width=3.0in,angle=0]{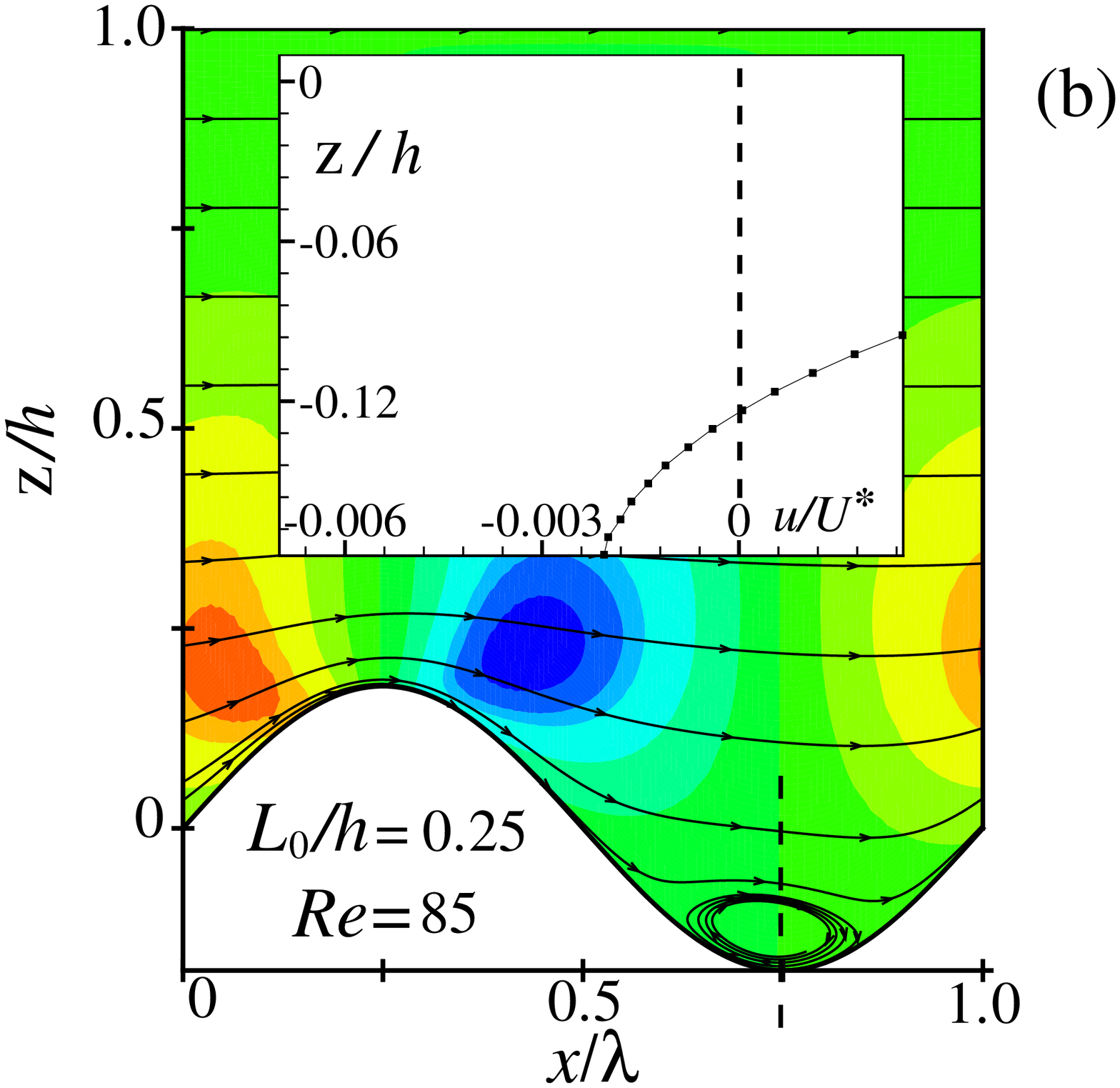}
\caption{(Color online) Pressure contours and streamlines extracted
from the solution of the Navier-Stokes equation for $ka=1.12$ and
$L_{0}=0$ (a) and $L_{0}/h=0.25$ (b). The value $P^{*}$ is the
maximum pressure located at $(x/\lambda,z/\lambda)\simeq(0.07,0.23)$
for $ka=1.12$, $L_{0}=0$, and $Re=4.0$. The dashed line inside the
valley indicates the cross-section used to compute the velocity
profiles shown in the inset.} \label{re_contour}
\end{center}
\end{figure}

The effect of the inertia term in the Navier-Stokes equation on the
shape of the wall shear stress and pressure profiles can be seen in
Fig.\,\ref{re_noslip}. The shear stress profiles above the peak are
not symmetric with respect to $x/\lambda=0.25$, which means that the
flow decelerates faster on the right side of the corrugation peak
than it accelerates on the left side. Also, the average adverse
pressure gradient on the left side of the peak is larger than its
value on the right side of the peak ($0.3\lesssim
x/\lambda\lesssim0.5$). By increasing the upper wall velocity, the
separation point moves further upstream than the attachment point
downstream. The formation of asymmetric vortex flow at finite
Reynolds numbers is consistent with previous findings for a flow in
an undulated tube~\cite{Nishimura84,Chow72,Sobey80,Leneweit99}. The
pressure contours and streamlines extracted from the NS equation
with the no-slip boundary condition are plotted in
Fig.\,\ref{re_contour}(a) for $ka=1.12$ and $Re=79$. The flow
streamlines in the valley indicate an asymmetric clockwise
circulation, which is larger than the flow circulation region shown
in Fig.\,\ref{contourslip}(a) for the Stokes case.

\begin{figure}
\begin{center}
\includegraphics[width=5in,angle=0]{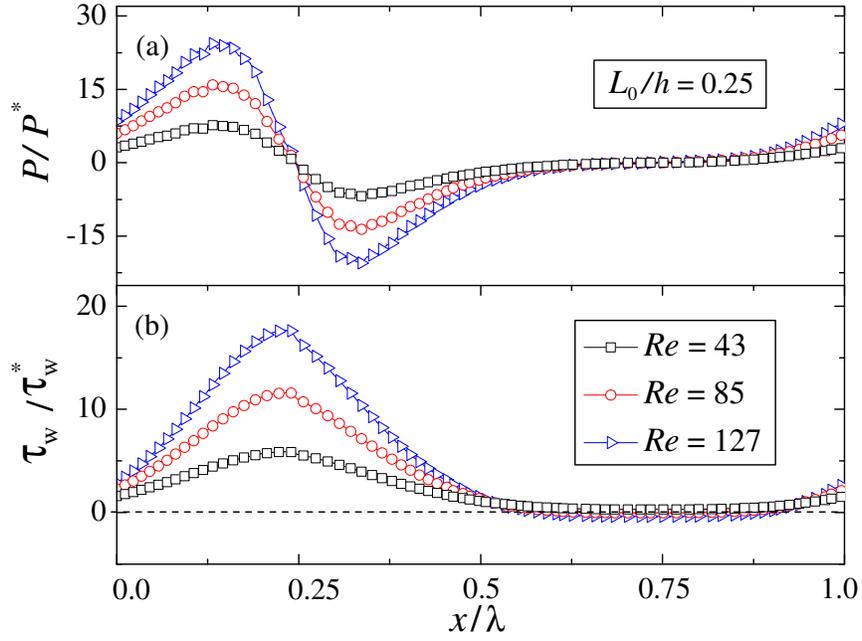}
\caption{(Color online) The normalized pressure (top) and shear
stress (bottom) along the lower wavy wall as a function of the
Reynolds number for $ka=1.12$ and $L_{0}/h=0.25$. The normalization
value $P^{*}$ is the maximum pressure located above the surface on
the left side of the peak $(x/\lambda,z/\lambda)\simeq(0.07,0.23)$
for $L_{0}=0$ and $Re=4.0$.} \label{re_slip}
\end{center}
\end{figure}

In the presence of the local slip condition along the lower
corrugated wall, the size of the vortex becomes smaller while the
flow structure remains asymmetric (see Fig.\,\ref{re_contour}). The
decrease of the vortex size is similar to the case of the Stokes
flow shown in Fig.\,\ref{contourslip} and can also be described in
terms of the pressure and shear stress along the lower wall. In
Fig.\,\ref{re_slip} the pressure and shear stress profiles are
plotted for the same values of the upper wall velocity as in
Fig.\,\ref{re_noslip} but with the slip boundary condition
($L_{0}/h=0.25$) along the lower wall. Note that the Reynolds
numbers in Fig.\,\ref{re_slip} are slightly larger than the values
reported in Fig.\,\ref{re_noslip} for the same $U^*$ because of the
larger effective slip length entering the definition of the Reynolds
number [see Eq.\,(\ref{Re_function})]. For each value of the upper
wall velocity, the adverse pressure gradient and the wall shear
stress on the right side of the peak are smaller than in the case of
the no-slip boundary condition, and, as a result, the vortex either
becomes smaller or vanishes (see Figs.\,\ref{re_noslip} and
\ref{re_slip}). As the Reynolds number increases, however, the
vortex forms and expands asymmetrically to fill the bottom of the
groove. The inset of the Fig.\,\ref{re_contour}(b) demonstrates that
the slip velocity at the bottom of the valley is negative while its
normal derivative $\partial u_{t}/\partial n$ is positive, in
contrast to the velocity profiles shown in Fig.\,\ref{contourslip}
for the same $ka=1.12$ and smaller slip lengths
$L_{0}/h\leqslant0.08$.

\begin{figure}
\begin{center}
\includegraphics[width=5in,angle=0]{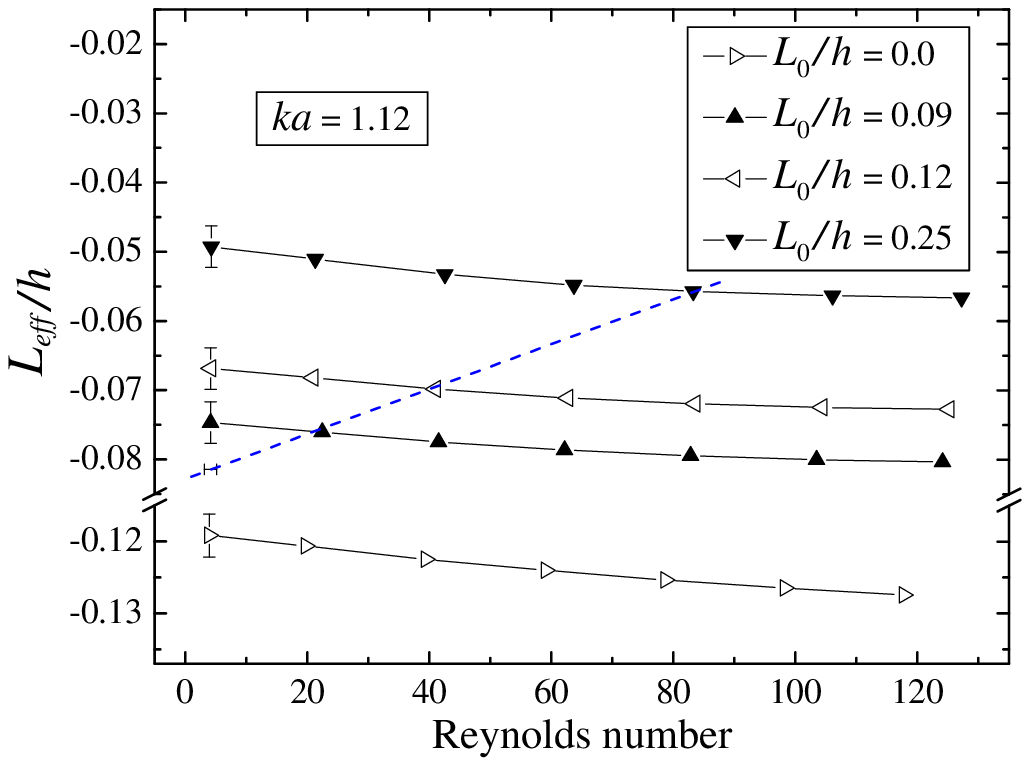}
\caption{(Color online) The effective slip length computed from the
solution of the Navier-Stokes equation as a function of the Reynolds
number for $ka=1.12$ and $L_{0}=0$ ($\vartriangleright$),
$L_{0}/h=0.09$ ($\blacktriangle$), $L_{0}/h=0.12$
($\vartriangleleft$), and $L_{0}/h=0.25$ ($\blacktriangledown$). The
dashed line indicates the upper bound of the region where a vortex
is present in the groove of a rough surface. The error bars
associated with the threshold values of the Reynolds number are
about the symbol size.} \label{re_leff}
\end{center}
\end{figure}

The effective slip length is plotted in Fig.\,\ref{re_leff} as a
function of the Reynolds number for the selected values of the
intrinsic slip length $L_0$ and $ka=1.12$. With increasing Reynolds
number, the flow streamlines move away from the lower boundary and
straighten out, the slope of the normalized velocity profiles
increases and the effective no-slip boundary plane is shifted into
the fluid domain. For $L_0/h\leqslant0.067$, the circulation is
always present in the valley and the flow streamlines in the bulk of
the fluid do not penetrate deep into the valley [e.g. see
Fig.\,\ref{re_contour}(b)]. For larger slip lengths $L_{0}/h>0.067$,
the flow streamlines show that there is no backflow at low $Re$, and
the vortex is formed at the bottom of the valley only at
sufficiently large Reynolds numbers indicated by the dashed line in
Fig.\,\ref{re_leff}. The numerical results obtained from the
solution of the Navier-Stokes equation demonstrate that the growth
or decay of the vortex as a function of the Reynolds number or the
intrinsic slip length is accompanied by the decrease or increase of
the effective slip length.

\section{Conclusions}

In this paper the effects of local slip boundary condition and the
Reynolds number on the flow structure near sinusoidally corrugated
surfaces and the effective slip length were investigated numerically
by solving the Stokes and Navier-Stokes equations. The effective
slip length was defined with respect to the mean height of the
surface roughness by extrapolating the linear part of the velocity
profile averaged over the corrugation period. In the case of the
Stokes flow with the local no-slip boundary condition, the effective
slip length decreases with increasing corrugation amplitude and the
vortex flow develops in the groove of the rough surface for
$ka\geqslant0.79$. In the presence of the local slip boundary
condition along the wavy wall, the effective slip length increases
and the size of the recirculation zone is reduced. The vortex
vanishes at sufficiently large values of the intrinsic slip length.
The analysis of the pressure and wall shear stress computed from the
Navier-Stokes equation shows that the asymmetric vortex flow
develops in the groove due to the inertia term even when the local
slip boundary condition is applied. The effective slip length
decreases with increasing Reynolds number. The numerical simulations
suggest that the variation of the vortex size as a function of
either the Reynolds number or the intrinsic slip length correlates
with the magnitude of the effective slip length.

\begin{acknowledgments}

Financial support from the Petroleum Research Fund of the American
Chemical Society is gratefully acknowledged. Computational work in
support of this research was performed at Michigan State
University's High Performance Computing Facility.

\end{acknowledgments}

\end{document}